\documentclass[twocolumn,prx,aps,amsmath,amssymb,longbibliography]{revtex4-1}

\usepackage{graphicx}
\usepackage{dcolumn}
\usepackage{bm}

\begin{document}

\title{Revealing universal Majorana fractionalization using differential shot
  noise and conductance in nonequilibrium states controlled by tunneling
  phases}

\author{Sergey Smirnov}
\affiliation{P. N. Lebedev Physical Institute of the Russian Academy of
  Sciences, 119991 Moscow, Russia}
\email{1) sergej.physik@gmail.com\\2)
  sergey.smirnov@physik.uni-regensburg.de\\3) ssmirnov@sci.lebedev.ru}

\date{\today}

\begin{abstract}
Universal fractionalization of quantum transport characteristics in
Majorana quantum dot devices is expected to emerge for well separated Majorana
bound states. We show that the Majorana universality of the differential shot
noise $\partial S^>/\partial V$ and conductance $\partial I/\partial V$ at low
bias voltages $V$ arises only in ideal setups with only one Majorana mode
entangled with the quantum dot. In realistic devices, where both Majorana
modes are entangled with the quantum dot, $\partial S^>/\partial V$ and
$\partial I/\partial V$ become very sensitive to tunneling phases and their
universal fractional values are hard to observe even for well separated
Majorana bound states. In contrast, as revealed here, the ratio
$(\partial S^>/\partial V)/(\partial I/\partial V)$ weakly depends on the
tunneling phases and is fractional when the Majorana bound states are well
separated or integer when they significantly overlap. Importantly, for very
large $V$ we demonstrate that this ratio becomes fully independent of the
tunneling phases and its universal fractional Majorana value may be observed
in state-of-the-art experiments.
\end{abstract}

\maketitle

\section{Introduction}\label{intro}
Reliable access to electronic degrees of freedom fractionalized by formation
of essentially nonlocal non-Abelian Majorana bound states (MBSs)
\cite{Kitaev_2001} in a topological superconductor (TS) is an appealing goal
which to some extent dates back to searching particles being their own
antiparticles \cite{Majorana_1937}. At present Majorana fractionalization is
of fundamental importance for exploring universal properties of condensed
matter systems \cite{Alicea_2012,Sato_2016,Lutchyn_2018,Valkov_2022}, where
MBSs appear in topologically nontrivial phases, and for practical
implementation of topologically protected qubits involved in anyonic quantum
computing \cite{Kitaev_2003}.

One way to access MBSs is via quantum transport characteristics sensitive to
fractionalizations of electronic degrees of freedom. Here average electric,
thermoelectric and heat currents
\cite{Liu_2011,Vernek_2014,Ramos-Andrade_2016,Liu_2017,Smirnov_2020a,Hong_2020,Chi_2020,Tang_2020,Zhang_2020,Chi_2020a,He_2021,Chi_2021,Wang_2021,Majek_2022,Bondyopadhaya_2022}
provide valuable characteristics of MBSs. Also dynamics of average
magnetizations \cite{Wrzesniewski_2021} provides an alternative approach to
study MBSs within quantum transport experiments. Qualitatively distinct and
much more detailed quantum transport information about MBSs is encoded in shot
and quantum noise which directly scans universal fractional character of
excitations in nonequilibrium states
\cite{Liu_2015,Liu_2015a,Haim_2015,Valentini_2016,Smirnov_2017,Smirnov_2018,Smirnov_2019,Smirnov_2019a,Feng_2022}.
In particular, Majorana shot noise taking into account continuum
quasiparticles in tunnel junctions \cite{Zazunov_2016}, giant Majorana shot
noise in topological trijunctions \cite{Jonckheere_2019} and full braiding
Majorana protocols obtained from weak measurements based on shot noise
statistics \cite{Manousakis_2020} demonstrate impressively diverse aspects of
Majorana physics beyond mean electric currents. Another opportunity which is
fundamentally different from quantum transport is to uniquely probe
fractionalization via quantum thermodynamic characteristics such as the
entropy. Indeed, well separated MBSs in nanoscopic setups result in fractional
plateaus \cite{Smirnov_2015,Sela_2019,Smirnov_2021,Smirnov_2021a,Ahari_2021}
of the entropy. Ongoing elaboration of experimental techniques and successful
measurements of the entropy
\cite{Hartman_2018,Kleeorin_2019,Pyurbeeva_2021,Child_2021,Child_2022a,Han_2022}
have established a convincing basis for a future thermodynamic access to
MBSs.

Due to well established technology to measure, {\it e.g.} electric currents,
quantum transport in nanoscopic setups is relatively simpler to perform than
quantum thermodynamics. In particular, measurements of mean electric currents
in setups presumably involving MBSs provide corresponding electric
conductances \cite{Mourik_2012}. Unfortunately, this most straightforward way
to probe MBSs is unreliable \cite{Yu_2021,Frolov_2021,Kejriwal_2022}. Thus
other approaches, based preferably on directly and simply measurable
quantities, are required to access MBSs beyond mean electric currents.

Here we demonstrate how one can access MBSs via directly measurable
observables such as differential shot noise and conductance in realistic
Majorana quantum dot (QD) devices. To avoid unreliable measurements
\cite{Yu_2021} in ideal setups, involving only one end of a TS, it is
necessary to probe MBSs simultaneously at both ends. Fractionalization in such
devices is driven by processes entangling a QD with both Majorana modes. Such
an entanglement may also be engineered for implementing Majorana qubits
\cite{Gau_2020,Gau_2020a} or even accidentally induced during a technological
process used to prepare a setup. We show that under such circumstances the
expected universal Majorana properties of the differential shot noise and
conductance are washed out by the Majorana tunneling phases even for well
separated MBSs. Thus each of these observables alone does not
straightforwardly indicate that in fact one deals with MBSs. This creates a
general problem because in majority of experimental setups precise values of
the tunneling phases are hard to extract. Remarkably, as we demonstrate, in
contrast to the differential shot noise and conductance, the ratio of these
two observables weakly depends on the tunneling phases. More importantly, we
find that it takes fractional values for well separated MBSs and becomes
integer when the MBSs merge into a single Dirac fermion. Thus this ratio
provides a straightforward access to fractionalization in Majorana QD
devices.

Furthermore, a detailed numerical analysis of the dependences of the
differential shot noise and conductance on the tunneling phases and bias
voltage reveals a fine structure of their resonances. Using these resonances
one can extract all the Majorana tunneling parameters including the tunneling
phases. The dependence on these parameters also suggests a way to shift the
Majorana resonances to higher bias voltages where it will be simpler to detect
them in experiments.
\begin{figure}
\includegraphics[width=8.0 cm]{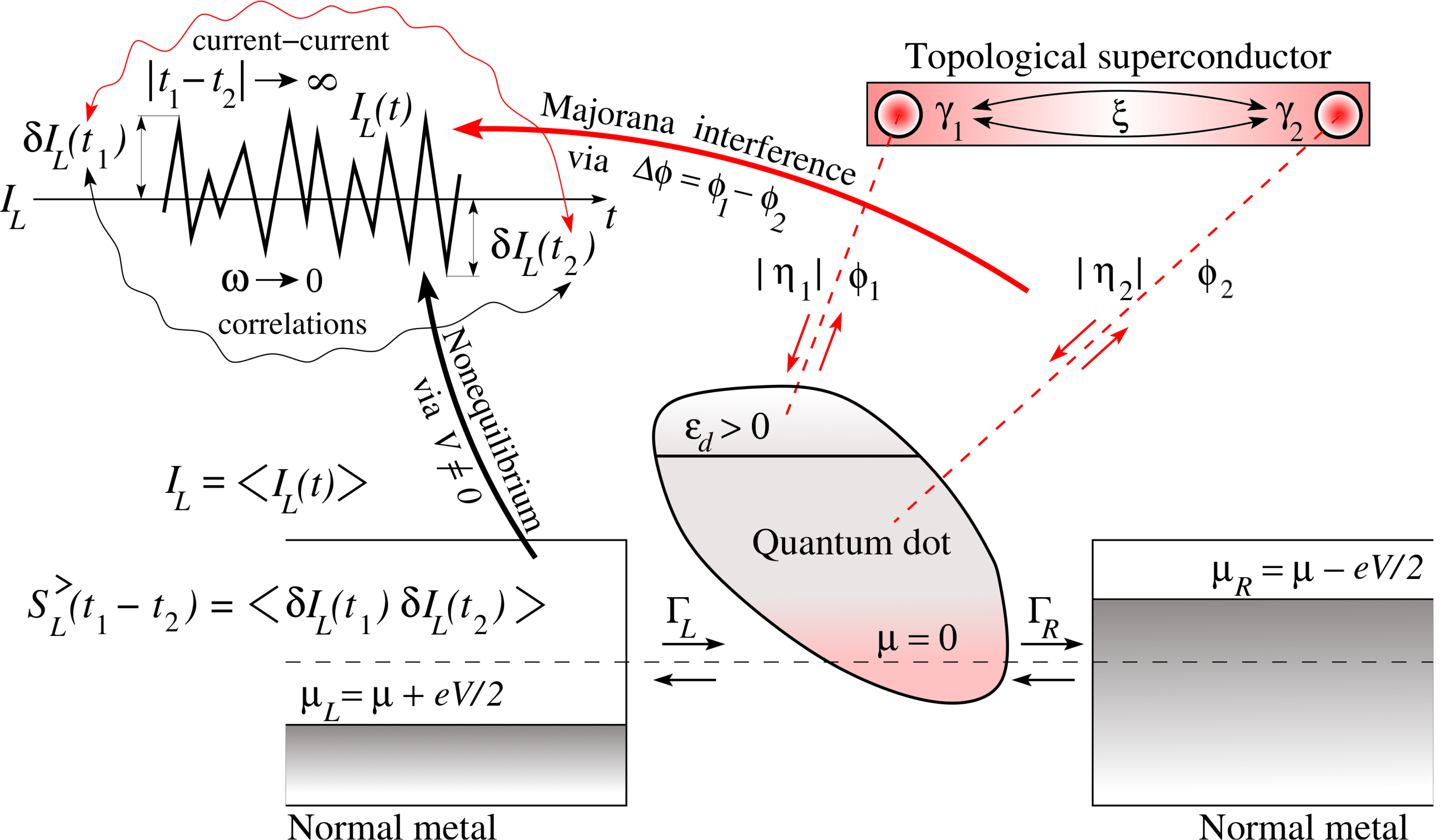}
\caption{\label{figure_1} A quantum device entangling both MBSs with external
  degrees of freedom. Due to this entanglement physical properties of the
  device, in particular quantum transport characteristics, such as its shot
  noise and conductance, strongly depend on the Majorana tunneling phases.}
\end{figure}

The paper is organized as follows. In Section \ref{mqdsdsnc} we present the
physical model of a feasible experimental setup which admits measurements of
quantum transport observables such as its differential shot noise and
conductance. Numerical results obtained for these observables are explored in
Section \ref{numres}. We conclude the paper with Section \ref{concl} where it
is shown that the Majorana universal fractionalization at high bias voltages
may be accessed in experiments performed at high temperatures achievable in
modern labs.
\section{Majorana quantum dot setup, its differential shot noise and
  conductance}\label{mqdsdsnc}
To quantitatively analyze how an entanglement of a QD with both Majorana modes
is revealed in fluctuations and the mean value of the electric current,
flowing through the QD, we consider the setup in
Fig. \ref{figure_1}. The schematic setup in Fig. \ref{figure_1} is fully
sufficient for our main goal, that is for the theoretical analysis presented
below. However, it is important to note that an experimental realization of
this theoretical model is feasible. For example, it can be implemented using
an InAs nanowire with an epitaxial Al layer grown on a part of the nanowire's
surface \cite{Deng_2016,Deng_2018}. Here, the Al shell is etched on one end of
the nanowire to prepare a bare InAs segment where afterwards a QD is
formed. Thus in this technological process the TS is located with respect to
the QD in an essentially asymmetric way: one end of the TS is much closer to
the QD than the other one. As a result, the first Majorana mode $\gamma_1$ is
coupled to the QD stronger than the second Majorana mode $\gamma_2$, that is
$|\eta_1|>|\eta_2|$ (see below for more details). The setup in
Ref. \cite{Deng_2018} may be adapted to our theoretical model by forming near
the QD two normal metallic contacts whose coupling strength $\Gamma$ (see
below for more details) may be varied by the voltage on gates located between
the contacts and QD. An alternative location of the TS with respect to the QD
is used in Ref. \cite{Liu_2011} and assumes a curved shape of the TS. In such
a curved setup both $|\eta_1|$ and $|\eta_2|$ are assumed to be controllable
and thus one can have both situations $|\eta_1|>|\eta_2|$ and
$|\eta_1|\approx|\eta_2|$. In particular, the regime $|\eta_1|>|\eta_2|$ is
used to implement driven dissipative Majorana qubits \cite{Gau_2020}. Below we
will assume $|\eta_1|>|\eta_2|$ since this situation is more relevant
technologically as well as for practical purposes. We also note that while in
the setup of Ref. \cite{Deng_2018} the phases $\phi_{1,2}$ may have arbitrary
fixed values induced during the technological process, in Ref. \cite{Liu_2011}
the phases $\phi_{1,2}$ are assumed to be externally controlled by a magnetic
flux. Such an external control of $\phi_{1,2}$ is also assumed for practical
implementations of Majorana qubits \cite{Gau_2020}. Our theoretical analysis
below is applicable to both of these situations and may be used to interpret
experiments where the phases $\phi_{1,2}$ are fixed or externally controlled.

To perform a theoretical analysis, we assume that the physical system includes
a QD,
\begin{equation}
  \hat{H}_d=\epsilon_d d^\dagger d,
  \label{ham_qd}
\end{equation}
with one nondegenerate energy level $\epsilon_d$, whose location may be tuned
by a gate voltage. Although here the QD is spinless and thus Kondo
correlations are absent, in the following we assume $\epsilon_d\geqslant 0$
(for example, empty dot, see also Ref. \cite{Vernek_2014}) in order to
describe a universal regime induced solely by MBSs, that is excluding even for
spin-degenerate QDs a possible interplay between universal Majorana
fractionalizations and the universality induced by Kondo correlations
\cite{Ralph_1994,Goldhaber-Gordon_1998,Glazman_1988,Meir_1993,Wingreen_1994,Smirnov_2011a,Smirnov_2011b}
which would have emerged in the spin-degenerate case for $\epsilon_d<0$. In
practice the spinless model in Eq. (\ref{ham_qd}) is realized, for example, in
strong magnetic fields used to bring the TS into its topological phase. Under
such circumstances interactions in the QD do not play a role and the
non-interacting spinless model in Eq. (\ref{ham_qd}) is an adequate tool to
analyze Majorana signatures \cite{Lopez_2014}. This has also been confirmed,
for example in Ref. \cite{Tijerina_2015}, using numerical renormalization
group calculations showing a transition of the linear conductance from the low
magnetic field plateau $3e^2/2h$ to the high magnetic field plateau $e^2/2h$
induced solely by the MBSs after the magnetic field has switched the QD into
the spinless regime and thus made it non-interacting via eliminating the Kondo
correlations. In spin-degenerate QDs interactions play an important role
\cite{Leijnse_2011,Lee_2013,Cheng_2014,Weymann_2017,Cifuentes_2019}. However,
numerical renormalization group calculations indicate (see, for example,
Ref. \cite{Silva_2020}) that even in interacting spin-degenerate QDs Majorana
induced fractionalizations may decouple from Kondo correlations. Thus
interacting spin-degenerate QDs coupled to MBSs may behave as their
non-interacting counterparts. In particular, the low-temperature entropies in
interacting and non-interacting spin-degenerate QDs coupled to MBSs are
identical and may be obtained using non-interacting spin-degenerate QDs
coupled to MBSs \cite{Silva_2020}. Thus even for spin-degenerate cases one may
consider non-interacting QDs coupled to MBSs to explore Majorana induced
properties at low temperatures, for example, Majorana shot noise as has been
done in Ref. \cite{Feng_2022}.

The setup also involves left ($L$) and right ($R$) contacts, which are normal
metals,
\begin{equation}
  \hat{H}_c=\sum_{l=L,R}\sum_k\epsilon_kc^\dagger_{lk}c_{lk},
  \label{ham_c}
\end{equation}
with continuum spectra approximated by a constant density of states,
$\nu(\epsilon)\approx\nu_c/2$, in the range of energies relevant for quantum
transport, and a grounded one-dimensional TS,
\begin{equation}
  \hat{H}_{ts}=i\xi\gamma_2\gamma_1/2,
  \label{ham_ts}
\end{equation}
whose low-energy properties are governed by MBSs,
$\gamma^\dagger_{1,2}=\gamma_{1,2}$, $\{\gamma_i,\gamma_j\}=2\delta_{ij}$,
localized at its ends and having a finite overlap, $\xi\neq 0$. The contacts'
Fermi-Dirac distributions,
\begin{equation}
  n_{L,R}(\epsilon)=\frac{1}{\exp\bigl[\frac{\epsilon-\mu_{L,R}}{k_\text{B}T}\bigr]+1},
  \label{FD_distr}
\end{equation}
are specified by the contacts' temperature $T$ and chemical potentials
$\mu_{L,R}=\mu\pm eV/2$, where $V$ is the bias voltage. The QD is coupled to
the contacts and TS via tunneling processes,
\begin{equation}
\begin{split}
&\hat{H}_{d\leftrightarrow c}=\sum_{l=L,R}\mathcal{T}_l\sum_kc^\dagger_{lk}d+\text{H.c.},\\
&\hat{H}_{d\leftrightarrow ts}=\eta^*_1d^\dagger\gamma_1+\eta^*_2d^\dagger\gamma_2+\text{H.c.},
\end{split}
\label{ham_tun}
\end{equation}
where $\eta_{1,2}=|\eta_{1,2}|\exp(i\phi_{1,2})$ and we assume
$|\eta_1|>|\eta_2|$. The interaction between the QD and contacts is
characterized by the tunneling energy $\Gamma\equiv\Gamma_L+\Gamma_R$,
$\Gamma_{L,R}\equiv\pi\nu_c|\mathcal{T}_{L,R}|^2$, and for simplicity we
assume $\Gamma_L=\Gamma_R$. When $V\neq 0$, different chemical potentials,
$\mu_L\neq\mu_R$, induce nonequilibrium which essentially determines the
electric current, in particular, its mean value and random deviations from
it. The MBSs $\gamma_1$ and $\gamma_2$ tunnel to the QD with different phases,
$\phi_1\neq\phi_2$, inducing on the QD an interference controlled by the phase
difference $\Delta\phi=\phi_1-\phi_2$ which also essentially determines the
mean value and fluctuations of the electric current.

Having the full Hamiltonian,
$\hat{H}=\hat{H}_d+\hat{H}_c+\hat{H}_{ts}+\hat{H}_{d\leftrightarrow c}+\hat{H}_{d\leftrightarrow ts}$,
one may consider various observables using, {\it e.g.}, the Keldysh field
integral formalism \cite{Altland_2010} particularly convenient to treat
stationary nonequilibrium. Within this formalism the second quantized
fermionic operators are replaced by the Grassmann fields, $\psi(t)$,
$\phi_{lk}(t)$, $\zeta(t)$, corresponding to the QD, contacts and TS,
respectively. The electric current operator in the $l$-th contact is expressed
through the Grassmann fields on the forward ($q=+$) or backward ($q=-$) branch
of the Keldysh contour,
\begin{equation}
  I_{lq}(t)=(ie/\hbar)\sum_k[\mathcal{T}_l\bar{\phi}_{lkq}(t)\psi_q(t)-\text{G.c.}],
  \label{el_curr_opr}
\end{equation}
where $\text{G.c.}$ denotes the Grassmann conjugation. The Hamiltonians in
Eqs. (\ref{ham_qd})-(\ref{ham_ts}) are replaced by the corresponding Keldysh
actions of the isolated QD, contacts and TS, $S_d$, $S_c$ and $S_{ts}$, which
are of conventional $2\times 2$ matrix form in the retarded-advanced
space. The Hamiltonians in Eq. (\ref{ham_tun}) are replaced by the following
actions:
\begin{equation}
  \begin{split}
  &S_{d\leftrightarrow c}=-\int_{-\infty}^\infty dt\sum_{l=L,R}\sum_k\{\mathcal{T}_l[\bar{\phi}_{lk+}(t)\psi_+(t)\\
    &-\bar{\phi}_{lk-}(t)\psi_-(t)]+\text{G.c.}\},\\
  &S_{d\leftrightarrow ts}=-\int_{-\infty}^\infty dt\{\eta_1^*[\bar{\psi}_+(t)\zeta_+(t)+\bar{\psi}_+(t)\bar{\zeta}_+(t)\\
      &-\bar{\psi}_-(t)\zeta_-(t)-\bar{\psi}_-(t)\bar{\zeta}_-(t)]+i\eta_2^*[\bar{\psi}_+(t)\zeta_+(t)\\
      &+\bar{\psi}_-(t)\bar{\zeta}_-(t)-\bar{\psi}_-(t)\zeta_-(t)-\bar{\psi}_+(t)\bar{\zeta}_+(t)]+\text{G.c.}\}.
  \end{split}
  \label{tun_act}
\end{equation}
To generate the mean electric current and various current-current correlators
one also adds the current action,
\begin{equation}
  S_I[J_{lq}(t)]=-\int_{-\infty}^\infty dt\sum_{l=L,R}\sum_{q=+,-}J_{lq}(t)I_{lq}(t),
  \label{scr_act}
\end{equation}
with the source fields $J_{lq}(t)$. The full Keldysh action,
$S_K[J_{lq}(t)]=S_d+S_c+S_{ts}+S_{d\leftrightarrow c}+S_{d\leftrightarrow ts}+S_I[J_{lq}(t)]$,
determines the Keldysh generating functional,
\begin{equation}
  Z[J_{lq}(t)]=\int\mathcal{D}[\bar{\psi}_q,\psi_q;\bar{\phi}_{lkq},\phi_{lkq};\bar{\zeta}_q,\zeta_q]e^{\frac{i}{\hbar}S_K[J_{lq}(t)]},
  \label{K_gen_func}
\end{equation}
from which one, {\it e.g.}, generates
\begin{equation}
  \begin{split}
  &\langle I_{lq}(t)\rangle_{S_K}=i\hbar\frac{\delta Z[J_{lq}(t)]}{\delta J_{lq}(t)}\biggl|_{J_{lq}(t)=0},\\
  &\langle I_{lq}(t)I_{l'q'}(t')\rangle_{S_K}=(i\hbar)^2\frac{\delta^2 Z[J_{lq}(t)]}{\delta J_{lq}(t)\delta J_{l'q'}(t')}\biggl|_{J_{lq}(t)=0},
  \end{split}
  \label{diff_gen_func}
\end{equation}
where $\langle\ldots\rangle_{S_K}$ means averaging at $J_{lq}(t)=0$ that is
with respect to the action $S_K=S_K[J_{lq}(t)=0]$,
\begin{equation}
  \langle\ldots\rangle_{S_K}=\int\mathcal{D}[\bar{\psi}_q,\psi_q;\bar{\phi}_{lkq},\phi_{lkq};\bar{\zeta}_q,\zeta_q]e^{\frac{i}{\hbar}S_K}[\dots].
  \label{avr_SK}
\end{equation}

In the following we will focus on the left contact, $l=L$, and explore the
mean electric current $I(V,\Delta\phi)=\langle I_{Lq}(t)\rangle_{S_K}$ and the
greater current-current correlator
$S^>(t,t';V,\Delta\phi)=\langle\delta I_{L-}(t)\delta I_{L+}(t')\rangle_{S_K}$,
$\delta I_{Lq}(t)=I_{Lq}(t)-I(V,\Delta\phi)$. The Fourier transform of the
correlator $S^>(t,t';V,\Delta\phi)=S^>(t-t';V,\Delta\phi)$,
\begin{equation}
  S^>(\omega;V,\Delta\phi)=\int_{-\infty}^\infty dt\,e^{i\omega t}S^>(t;V,\Delta\phi),
  \label{Ft_gt_curr_curr}
\end{equation}
is an important quantity because at $\omega=0$ it specifies the shot noise
$S^>(V,\Delta\phi)\equiv S^>(\omega=0;V,\Delta\phi)$.

In experiments one directly measures the differential shot noise and
conductance, $\partial S^>(V,\Delta\phi)/\partial V$,
$\partial I(V,\Delta\phi)/\partial V$, and below we focus on these partial
derivatives. We obtain $S^>(V,\Delta\phi)$ and $I(V,\Delta\phi)$ via numerical
integrations and then perform numerical differentiations. The regime where
$|\eta_1|$ prevails,
$|\eta_1|>\text{max}\{|\epsilon_d|,|eV|,k_\text{B}T,\Gamma,|\eta_2|,\xi\}$, is
of particular interest for observing Majorana universality in experiments and
below we explore it explicitly. As discussed above, the physical setup assumes
that $\Gamma$ may be enhanced or suppressed by the voltage on gates located
between the QD and contacts. Our motivation for suppressing $\Gamma$ below
$|\eta_1|$ is that for $\Gamma<|\eta_1|$ the properties of the QD will be
dominated by Majorana fractional character in a wide energy range. Physically
this is expected because the tunneling of the MBSs into the QD is much
stronger than the tunneling from the contacts whose non-fractional degrees of
freedom could significantly wash out various Majorana fractionalizations in
the opposite regime, that is for $\Gamma>|\eta_1|$. We also note that to get
the numerical results presented below we specify all the energies in units of
$\Gamma$. In particular, the bias voltage $|eV|$ is varied in a wide range,
from small to large values in units of $\Gamma$. However, since $\Gamma$ is
made very small using the gates mentioned above, $|eV|$ will also be small
even if it takes large values in units of $\Gamma$. This means that in our
numerical results the bias voltage always remains small enough so that other
levels of the QD are not relevant and our theoretical model provides a
reasonable description of experiments.
\begin{figure}
\includegraphics[width=8.0 cm]{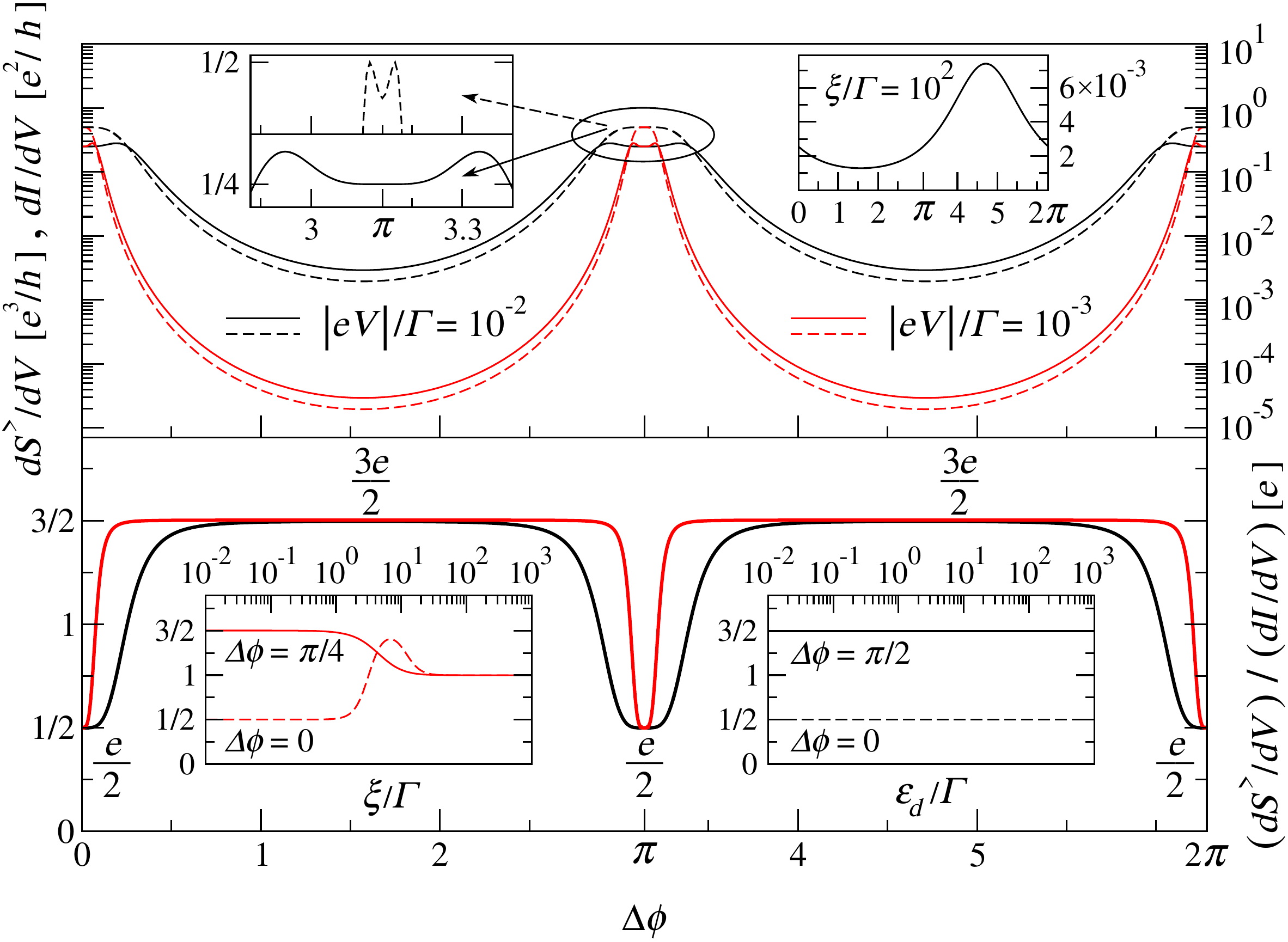}
\caption{\label{figure_2} Upper panel: Differential shot noise
  $\partial S^>/\partial V$ (solid lines) and differential conductance
  $\partial I/\partial V$ (dashed lines) as functions of the tunneling phase
  difference $\Delta\phi$ for small bias voltages, $|eV|/\Gamma\ll 1$. The
  black curves correspond to the case $|eV|/\Gamma=10^{-2}$ while the red
  curves correspond to the case $|eV|/\Gamma=10^{-3}$. The other parameters
  are the same for the two cases: $\epsilon_d/\Gamma=10$,
  $k_\text{B}T/\Gamma=10^{-6}$, $|\eta_1|/\Gamma=10^3$,
  $|\eta_2|/\Gamma=10^{-1}$, $\xi/\Gamma=10^{-4}$. Lower panel: Black and red curves show the ratio
  $(\partial S^>/\partial V)/(\partial I/\partial V)$ obtained from the
  corresponding curves in the upper panel.}
\end{figure}
\section{Numerical results}\label{numres}
In the upper panel of Fig. \ref{figure_2} we show $\partial S^>/\partial V$
and $\partial I/\partial V$ as functions of $\Delta\phi$ for
$|eV|/\Gamma\ll 1$. Both quantities exhibit resonances around
$\Delta\phi_\text{res}=0,\,\pi,\,2\pi$ where they reach their universal
fractional values, $\partial S^>/\partial V=e^3/4h$ and
$\partial I/\partial V=e^2/2h$. However, due to strong dependences of
$\partial S^>/\partial V$ and $\partial I/\partial V$ on $\Delta\phi$ they
quickly deviate from these fractional values which, thus, can be observed only
in very narrow ranges of $\Delta\phi$. The left inset shows the fine
structures of the resonances of $\partial S^>/\partial V$ and
$\partial I/\partial V$ in the vicinity of $\Delta\phi=\pi$. As can be seen,
these resonances are split into two peaks. We find numerically that the
distances $\Delta\phi_{I,\text{max}}$ and $\Delta\phi_{S,\text{max}}$ between
the peaks of $\partial I/\partial V$ and $\partial S^>/\partial V$,
respectively, depend on both $|\eta_2|$ and $|eV|$,
\begin{equation}
  \Delta\phi_{I,\text{max}}=\frac{1}{2}\frac{|eV|}{|\eta_2|},\quad\Delta\phi_{S,\text{max}}\sim\frac{\sqrt{\Gamma|eV|}}{|\eta_2|}.
  \label{dist_max_I_S}
\end{equation}
These expressions are valid only when $|eV|\gg k_\text{B}T$. For
$|eV|\ll k_\text{B}T$ the resonance of the linear conductance also splits into
two resonances with the distance between them depending on $|\epsilon_d|$ as
also confirmed by the entropy analysis \cite{Smirnov_2021}. In particular, for
$\epsilon_d=0$ it is equal to $\pi$ \cite{Liu_2011}. The right inset shows
$\partial S^>/\partial V$ and $\partial I/\partial V$ for $\xi/\Gamma\gg 1$
when the MBSs strongly overlap and form a single Dirac fermion. Specifically,
in the inset $\xi/\Gamma=10^2$. The two physical quantities,
$\partial S^>/\partial V$ and $\partial I/\partial V$, are strongly suppressed
and coincide with each other leading to a single curve in the inset. Moreover,
as the inset explicitly demonstrates, the strong overlap of the MBSs doubles
the period of $\partial S^>/\partial V$ and $\partial I/\partial V$ from $\pi$
to $2\pi$. The lower panel shows the ratio
\begin{figure}
\includegraphics[width=8.0 cm]{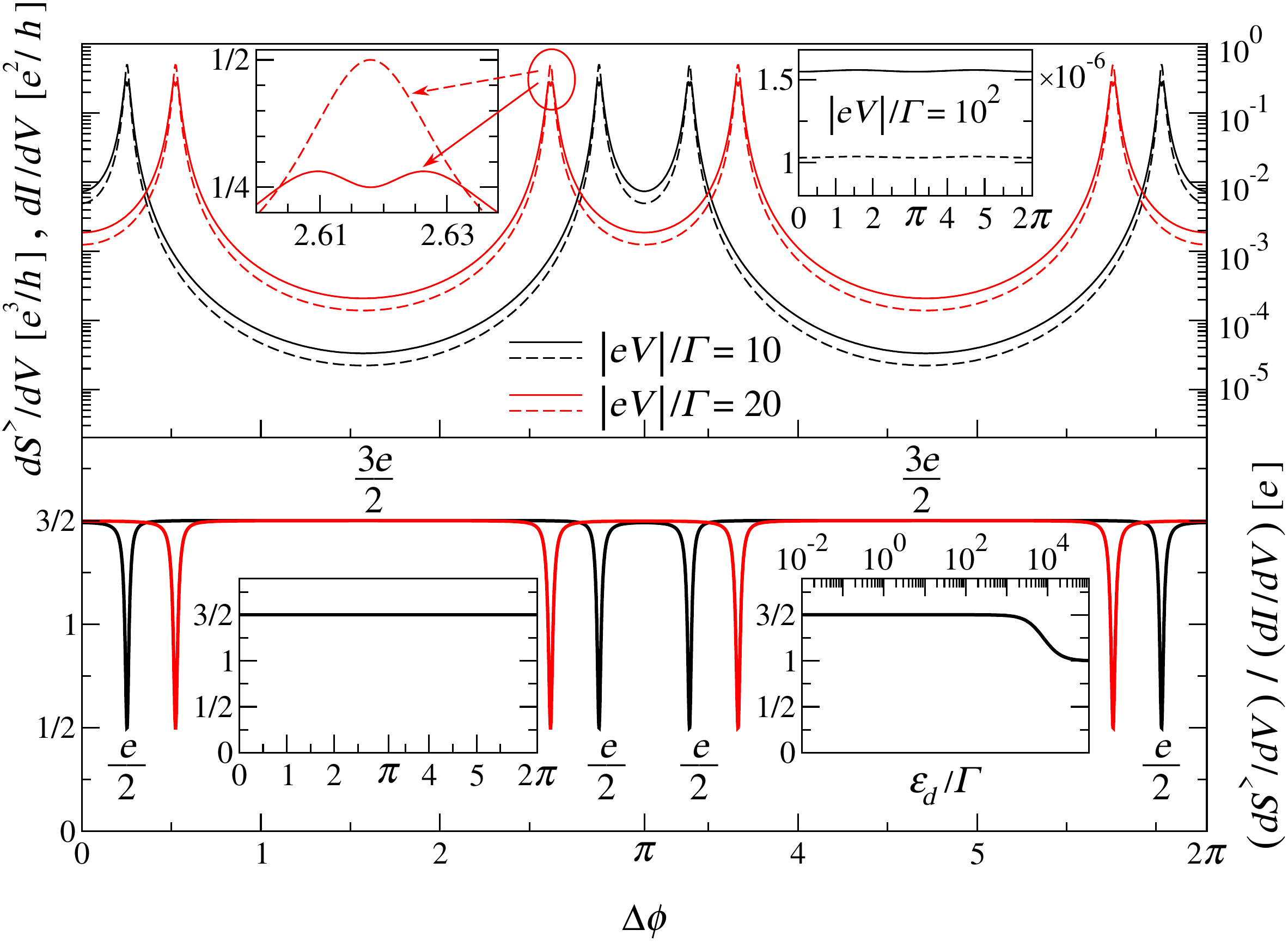}
\caption{\label{figure_3} Upper panel: Differential shot noise
  $\partial S^>/\partial V$ (solid lines) and differential conductance
  $\partial I/\partial V$ (dashed lines) as functions of the tunneling phase
  difference $\Delta\phi$ for large bias voltages, $|eV|/\Gamma\sim10$. The
  black curves correspond to the case $|eV|/\Gamma=10$ while the red curves
  correspond to the case $|eV|/\Gamma=20$. The other parameters are the same
  for the two cases: $\epsilon_d/\Gamma=10$, $k_\text{B}T/\Gamma=10^{-6}$,
  $|\eta_1|/\Gamma=10^3$, $|\eta_2|/\Gamma=10$, $\xi/\Gamma=10^{-4}$. Lower
  panel: Black and red curves show the ratio
  $(\partial S^>/\partial V)/(\partial I/\partial V)$ obtained from the
  corresponding curves in the upper panel.}
\end{figure}
$(\partial S^>/\partial V)/(\partial I/\partial V)$ corresponding to the black 
and red curves in the upper panel. In contrast to $\partial S^>/\partial V$
and $\partial I/\partial V$ their ratio is almost independent of
$\Delta\phi$. Indeed, it has wide plateaus, on which it is equal to $3e/2$,
and narrow antiresonances, corresponding to the resonances in the upper
panel. The minimum of these antiresonances is also fractional, equal to $e/2$,
while their width is determined by both the bias voltage $|eV|$ and tunneling
amplitude $|\eta_2|$. The left inset shows the ratio
$(\partial S^>/\partial V)/(\partial I/\partial V)$ as a function of $\xi$ for
the case $|eV|/\Gamma=10^{-3}$ and $\Delta\phi=0$ (dashed curve),
$\Delta\phi=\pi/4$ (solid curve). We see that both fractional values, $3e/2$
and $e/2$, are observed for well separated MBSs, when $\xi/\Gamma\ll 1$. For
strongly overlapping MBSs, when $\xi/\Gamma\gg 1$, the ratio takes the integer
Dirac value, $(\partial S^>/\partial V)/(\partial I/\partial V)=e$, and
becomes independent of $\Delta\phi$. The right inset shows for the case
$|eV|/\Gamma=10^{-2}$ the universality of the two nontrivial Majorana
fractional values, $3e/2$ (solid curve, $\Delta\phi=\pi/2$) and $e/2$ (dashed
curve, $\Delta\phi=0$), that is their independence of the gate voltage
controlling the value of $\epsilon_d$.

The differential shot noise and conductance for large bias voltages are shown
as functions of $\Delta\phi$ in the upper panel of Fig. \ref{figure_3}. As in
Fig. \ref{figure_2}, the curves exhibit resonances with
$\partial S^>/\partial V=e^3/4h$ and $\partial I/\partial V=e^2/2h$. However,
now $\Delta\phi_\text{res}\neq 0,\,\pi,\,2\pi$. The new values of
$\Delta\phi_\text{res}$ are determined by both $|eV|$ and $|\eta_2|$. Here,
the value of $|\eta_2|$ is increased in comparison with Fig. \ref{figure_2} in
order to clearly show that the new locations of the resonances of both
$\partial S^>/\partial V$ and $\partial I/\partial V$ are determined by
$\Delta\phi_{I,\text{max}}$ (and not by $\Delta\phi_{S,\text{max}}$) in
Eq. (\ref{dist_max_I_S}). The left inset shows that at large bias voltages
there happens a qualitative change in the structure of the resonances in
comparison with small bias voltages. Indeed, as can be seen in this inset, the
resonance of $\partial I/\partial V$ does not have any fine structure anymore
and is characterized by a single peak. In contrast, the resonance of $\partial
S^>/\partial V$ has a fine structure: it is split into two peaks. From our
numerical calculations we find that the distance between these two peaks
depends on $|\eta_2|$ and almost independent of $|eV|$,
\begin{equation}
  \Delta\phi_{S,\text{max}}\sim\frac{\Gamma}{|\eta_2|}.
  \label{dist_max_S_large_V}
\end{equation}
The right inset shows $\partial S^>/\partial V$ (solid curve) and
$\partial I/\partial V$ (dashed curve) as functions of $\Delta\phi$ for very
large bias voltage, $|eV|/\Gamma=10^2$. As can be seen, at very large bias
voltages both the differential shot noise and conductance are strongly
suppressed, $\partial S^>/\partial V\ll e^3/4h$,
$\partial I/\partial V\ll e^2/2h$, and vary relatively weakly as functions of
$\Delta\phi$. The curves in the lower panel demonstrate the ratio
$(\partial S^>/\partial V)/(\partial I/\partial V)$ corresponding to the two
bias voltages in the upper panel. As in Fig. \ref{figure_2}, whereas
$\partial S^>/\partial V$ and $\partial I/\partial V$ change relatively
strongly (several orders of magnitude), their ratio exhibits weak dependence
on $\Delta\phi$ characterized by wide plateaus separated by narrow
antiresonances. On the plateaus and at the minima of the antiresonances the
ratio is fractional, equal to $3e/2$ and $e/2$, respectively. The left inset
shows the ratio $(\partial S^>/\partial V)/(\partial I/\partial V)$ obtained
from the curves in the right inset of the upper panel. We see that although at
very large bias voltages ($|eV|/\Gamma=10^2$) both $\partial S^>/\partial V$
and $\partial I/\partial V$ are strongly suppressed, their ratio takes the
Majorana fractional value $3e/2$ in the whole range of $\Delta\phi$. The right
inset demonstrates the universality (independence of $\epsilon_d$) of the
Majorana fractional value $3e/2$ at very large bias voltages
($|eV|/\Gamma=10^2$). The curve in this inset is also independent of
$\Delta\phi$. As expected, the Majorana universality takes place only at
$|\epsilon_d|\leqslant|\eta_1|$.
\begin{figure}
\includegraphics[width=8.0 cm]{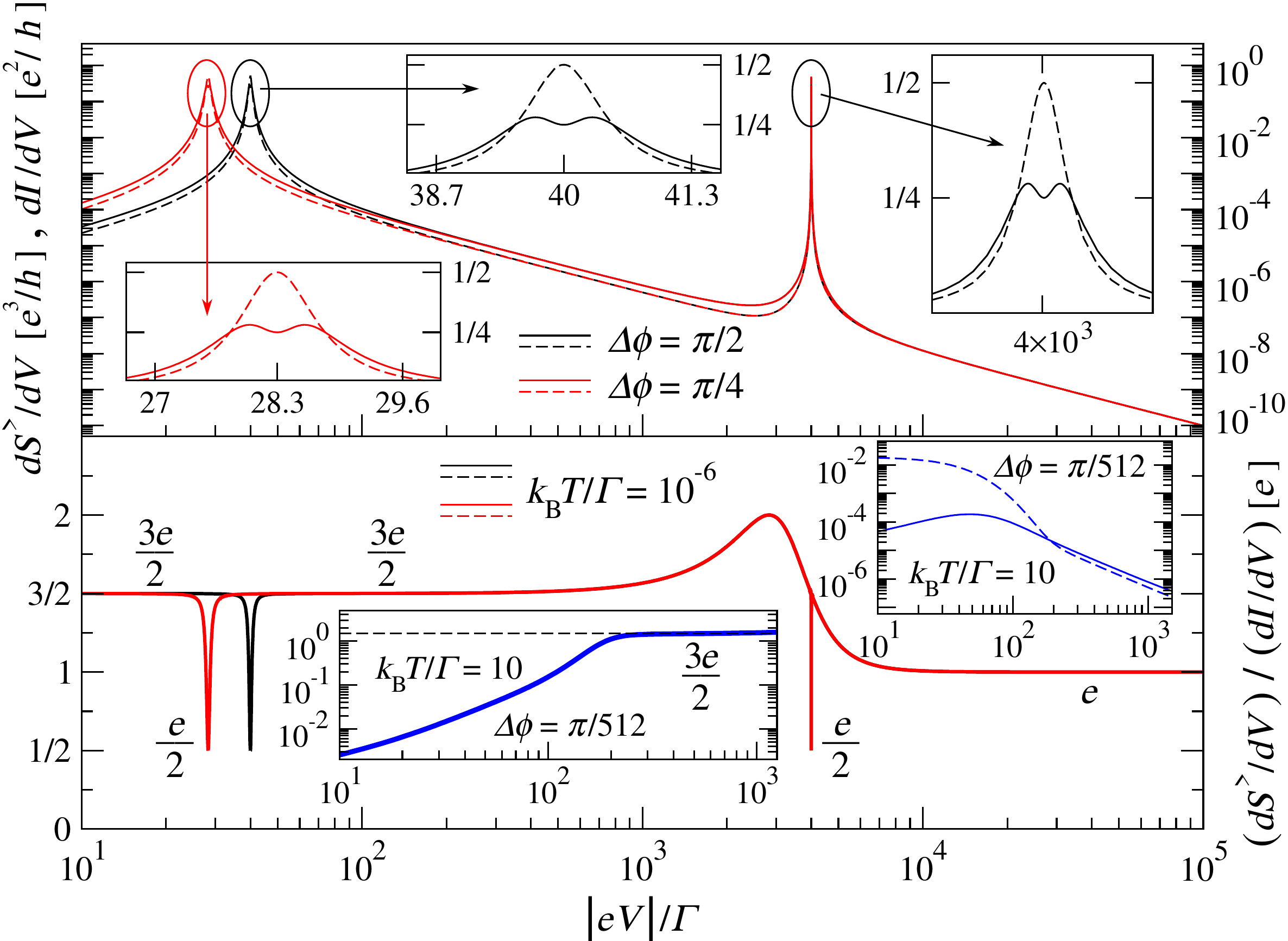}
\caption{\label{figure_4} Upper panel: Differential shot noise
  $\partial S^>/\partial V$ (solid lines) and differential conductance
  $\partial I/\partial V$ (dashed lines) as functions of the bias voltage
  $|eV|$ for two values of the tunneling phase difference $\Delta\phi$. The
  black curves correspond to the case $\Delta\phi=\pi/2$ while the red curves
  correspond to the case $\Delta\phi=\pi/4$. The other parameters are the same
  for the two cases: $\epsilon_d/\Gamma=10$, $k_\text{B}T/\Gamma=10^{-6}$,
  $|\eta_1|/\Gamma=10^3$, $|\eta_2|/\Gamma=10$, $\xi/\Gamma=10^{-4}$. Lower
  panel: Black and red curves show the ratio
  $(\partial S^>/\partial V)/(\partial I/\partial V)$ obtained from the
  corresponding curves in the upper panel.}
\end{figure}

The upper panel of Fig. \ref{figure_4} shows $\partial S^>/\partial V$ and
$\partial I/\partial V$ as functions of $V$ for two different values of
$\Delta\phi$. Both $\partial S^>/\partial V$ and $\partial I/\partial V$
exhibit two resonances at $|eV_\text{res,1}|$ and $|eV_\text{res,2}|$ where
they reach their universal fractional values, $\partial S^>/\partial V=e^3/4h$
and $\partial I/\partial V=e^2/2h$. We find numerically that $|eV_\text{res,1}|$
depends only on $|\eta_1|$ whereas $|eV_\text{res,2}|$ depends on both
$|\eta_2|$ and $\Delta\phi$,
\begin{equation}
  |eV_\text{res,1}|=4|\eta_1|,\quad|eV_\text{res,2}|=4|\eta_2|\sin(\Delta\phi).
  \label{res_I_S_bias}
\end{equation}
Away from these resonances both $\partial S^>/\partial V$ and
$\partial I/\partial V$ quickly deviate from their universal fractional
values. The left and middle insets show the detailed profiles of the
resonances at $|eV_\text{res,2}|$ for $\Delta\phi=\pi/4$ and
$\Delta\phi=\pi/2$, respectively, while the right inset shows the detailed
profile of the resonances at $|eV_\text{res,1}|$. These three insets all
demonstrate that the resonances of $\partial I/\partial V$ have a simple
structure of a single peak while the resonances of $\partial S^>/\partial V$
have a fine structure composed of two peaks. From numerical calculations we
find that the distance between these two peaks is approximately equal to
$\Gamma$ and does not depend on the other parameters. In the lower panel we
show the ratio $(\partial S^>/\partial V)/(\partial I/\partial V)$
corresponding to the two curves in the upper panel. Note, while the variations
of the $\partial S^>/\partial V$ and $\partial I/\partial V$ reach several
orders of magnitude, their ratio changes relatively weakly and shows a number
of characteristic features. Specifically, in the Majorana regime, that is for
$|eV|<|\eta_1|$, this ratio is characterized by plateaus separated by an
antiresonance located at $|eV_\text{res,2}|$. At these plateaus the ratio is
equal to $3e/2$ and at the minimum of the antiresonance it is equal to
$e/2$. Also at $|eV_\text{res,1}|$ there exists an antiresonance with the same
minimum $e/2$. For very large bias voltages, $|eV|\gg|\eta_1|$, when the
Majorana tunneling is ineffective, the two Majorana modes behave as a single
Dirac fermion and there forms a trivial plateau with the Dirac value
$(\partial S^>/\partial V)/(\partial I/\partial V)=e$.
\section{Conclusion}\label{concl}
We have demonstrated how to reveal Majorana fractionalization using quantum
transport characteristics such as differential shot noise and conductance in
Majorana QD setups where both Majorana modes are entangled with a strongly
nonequilibrium QD. Our numerical results show that the tunneling phases wash
out the Majorana universal values of each of these two observables. At the
same time it has been found that the ratio of the differential shot noise and
conductance weakly depends on the tunneling phases and provides a reliable
access to Majorana fractionalization in realistic strongly nonequilibrium
setups. In particular, it has been explicitly demonstrated that this ratio
takes fractional values when MBSs in a TS are well separated whereas the
ratio becomes integer for strongly overlapping MBSs merging into a single
Dirac fermion. Additionally, we have demonstrated the universality of the
Majorana fractionalization that is its independence not only of the tunneling
phase difference $\Delta\phi$ but also its independence of the QD gate voltage
$\epsilon_d$ and bias voltage $V$. Regarding the behavior of the differential
shot noise and conductance we have found that each of these observables has a
resonant structure in its dependence on both the tunneling phase difference
and bias voltage. Fine structures of these resonances have been explicitly
shown and explored.

Let us in conclusion probe the Majorana universality at high temperatures
achievable in state-of-the-art experiments. The insets in the lower panel of
Fig. \ref{figure_4} illustrate results obtained for
$k_\text{B}T/\Gamma=10$. At both of the insets $\Delta\phi=\pi/512$ and the
other parameters remain unchanged. The two curves in the right inset show
$\partial S^>/\partial V$ (solid) and $\partial I/\partial V$ (dashed). The
curve in the left inset shows the ratio
$(\partial S^>/\partial V)/(\partial I/\partial V)$ obtained from the two
curves in the right inset. As one can see in the left inset, high temperatures
destroy the low-energy part of the universal fractional Majorana
plateau. However, for bias voltages in the range $k_\text{B}T\ll|eV|<|\eta_1|$
the high-energy part of this plateau is still present and provides an
experimental access to the universal fractional Majorana ratio $3e/2$.

Finally, we note that our results enable to extract all the Majorana tunneling
parameters. Indeed, according to Eq. (\ref{dist_max_I_S}), one obtains
$|\eta_2|$ from measurements of $\Delta\phi_{I,\text{max}}$ since the bias
voltage $V$ is known in experiments. Measuring then $|eV_\text{res,1}|$ and
$|eV_\text{res,2}|$ one gets $|\eta_1|$ and $\Delta\phi$ using
Eq. (\ref{res_I_S_bias}). Besides, the expression for $|eV_\text{res,2}|$ in
Eq. (\ref{res_I_S_bias}) suggests that one may increase $|\eta_2|$ as
necessary to shift the corresponding Majorana resonances of
$\partial S^>/\partial V$ and $\partial I/\partial V$ to higher bias voltages
where it is easier to measure them at higher temperatures.
\section*{Acknowledgments}
The author thanks Reinhold Egger, Andreas K. H{\"u}ttel and Wataru Izumida for
useful discussions.

\end{document}